\documentclass[%
 reprint,
superscriptaddress,
 amsmath,amssymb,
 aps,
groupedaddress,
]{revtex4-1}
\usepackage{graphicx}
\usepackage{amsmath}
\usepackage{float}
\usepackage{comment}
\usepackage{amsmath}
\usepackage{amssymb}
\usepackage{mathrsfs}
\usepackage[below]{placeins}
\usepackage{fancyhdr}
\usepackage{soul,color}
\begin{document}


\title[\texttt{achemso} demonstration]
{Probing the phase diagram of cuprates with YBa$_2$Cu$_3$O$_{7-\delta}$ thin films and nanowires}


\author{Riccardo Arpaia}
\author{Eric Andersson}
\author{Edoardo Trabaldo}
\author{Thilo Bauch}
\author{Floriana Lombardi}
\email{floriana.lombardi@chalmers.se}

\affiliation{Quantum Device Physics Laboratory, Department of Microtechnology and Nanoscience, Chalmers University of Technology, SE-41296 G\"{o}teborg, Sweden}

\date{\today}
\begin{abstract}
We have grown and characterized 30 nm thick YBa$_2$Cu$_3$O$_{7-\delta}$ (YBCO) films, deposited by pulsed laser deposition on both MgO (110) and SrTiO$_3$ (001) substrates, which induce opposite strain to the superconducting layer. By carefully tuning the in-situ post-annealing oxygen pressure, we achieved, in a reproducible way, films at different oxygen doping, spanning from the slightly overdoped down to the strongly underdoped region of the phase diagram. The transport properties of the films, investigated through resistance versus temperature measurements, are in perfect qualitative agreement with single crystals. Starting from these films, we have also successfully fabricated nanowires with widths down to 65 nm, at different oxygen doping. The nanostructures exhibit characteristic temperatures (as the critical temperature $T_{\mathrm{c}}$ and the pseudogap temperature $T^*$) similar to those of the as-grown films and carry critical current densities $J_{\mathrm{c}}$ close to the critical      depairing value, limited by vortex entry. This implies that the superconducting and the normal state properties of underdoped YBCO are preserved in our films, and they can be studied as a function of the dimensionality of the system, down to the nanoscale.
\end{abstract}


\maketitle

\section{Introduction}

The microscopic origin of the superconducting phenomenon in High critical Temperature Superconductors (HTSs) is intimately related to the nature of the highly unconventional normal state, which is described by a very complex phase diagram. The still elusive pseudogap state, in the underdoped region of the cuprates, hosts several nanoscale orders, as the newly emerged charge density waves (CDW) \cite{ghiringhelli2012long, chang2012direct, da2015charge} and the electronic nematicity \cite{cyr2015two, sato2017thermodynamic}. These two orders break the spontaneous rotational and/or translational symmetry of the crystal. The electronic nematicity causes an anisotropy of the electron transport, which adds to the transport anisotropy due to the orthorhombicity of the crystals \cite{ando2002electrical}. Remarkably, when observed in thin films, it strongly depends on the strain induced by the substrate to the thin film \cite{wu2017spontaneous}.  The strain control in thin films can therefore be instrumental to understand the entanglement between various orders. This requires the growth of thin films, as a function of the oxygen doping, which can well reproduce the main features of the HTS phase diagram.



However, up to now in literature a complete phase diagram is available only for single crystals \cite{ando2004electronic, barivsic2013universal, hucker2014competing}. For thin films the strain induced by the substrate has been reported to affect the critical temperature $T_{\mathrm{c}}$. 
For instance, for the La$_{2-x}$Sr$_x$CuO$_4$ and La$_{2-x}$Ba$_x$CuO$_4$ compounds a compressive strain causes the complete disappearance of the $1/8$ anomaly, corresponding to the $T_{\mathrm{c}}$ suppression around the doping $p = 0.125$ (where $p$ represents the number of holes for planar copper atom). It also leads to a  change in the temperature dependence of the resistance above $T_{\mathrm{c}}$ with respect to single crystals \cite{sato20002, sato2000absence}. Tensile strain appears instead to be responsible for a strong reduction of $T_{\mathrm{c}}$ at any doping, together with a weakening of the $1/8$ anomaly \cite{sato2001influence}. 
Similarly,  works on YBa$_2$Cu$_3$O$_{7-\delta}$  (YBCO) thin films are rather limited. The presence of the peculiar depression of $T_{\mathrm{c}}$ at $p = 1/8$ is sometimes reported \cite{PhysRevB.48.7554, wu1998preparation}. However structural disorder and strain, preventing oxygen ordering, have more often led to the disappearance of this feature \cite{osquiguil1992controlled, macmanus1994studies, tolpygo1996universal, farnan2000fabrication, farnan2001doping}. The reproducible observation of the $1/8$ anomaly in the phase diagram of thin films is therefore the first step one needs to take to disclose the intertwining of different orders by using nanoscale devices.

Few works are devoted to the fabrication of underdoped nanostructures. Bonetti et al \cite{Bonetti2008} reports on the fabrication of underdoped YBCO nanowires, with widths down to 200 nm. The $T_{\mathrm{c}}$ of the nanowires is 5-10 K lower than that of the unpatterned films, which is already a sign of the degradation of the superconducting properties. Moreover nanowires, patterned on underdoped films having $T_{\mathrm{c}}$ lower than 75 K, are not superconducting. Carillo et al \cite{carillo2012coherent} show the realization of Nd$_{1.2}$Ba$_{1.8}$Cu$_3$O$_{\delta}$ nanodevices, starting from films having various $T_{\mathrm{c}}$, from the maximum of 65 K down to the non-superconducting state.  The $T_{\mathrm{c}}$ of sub-micron wires decreases dramatically - by tens of Kelvin - when reducing the width. Structures narrower than 200 nm are not anymore superconducting. Consequently, it is very hard to distinguish and possibly separate change in doping level from disorder, which are both sources of reduction of $T_{\mathrm{c}}$ in nanowires. 

So far underdoped HTS cuprates have never been studied at the nanoscale, on dimensions comparable with the CDW correlation length $\xi_{\mathrm{CDW}}$ (of the order of few nanometers at the $1/8$ doping).

In this paper we present the transport characterization of YBCO thin films, covering the whole underdoped region of the  $T$($p$) phase diagram. We have grown the films on two substrates, (110) MgO and (001) STO, having completely different matching with the YBCO layer, to study the effect of the strain on the electronic properties.  
The films, independently of the used substrate, reproduce a phase diagram very similar to that of single crystals, including the anomalies associated with CDW order at the $1/8$ doping. The films, protected by a Au capping layer, have been patterned into nanowires, having widths down to 65 nm. At any oxygen doping, the nanowires have a critical current density $J_{\mathrm{c}}$ very close to the theoretical depairing limit, and a critical temperature $T_{\mathrm{c}}$ very close to the as-grown films. These nanowires are    therefore an ideal platform to study the rich and complex physics, characterizing the underdoped region of the YBCO phase diagram, at the nanoscale. 

\section{Film deposition}

We have deposited 30 nm thick YBCO films on (110) oriented MgO and (001) oriented SrTiO$_3$ (STO) substrates by pulsed laser deposition (heater temperature 760 $^\circ$C, oxygen pressure 0.7 mbar, energy density 1.5 J/cm$^2$). In previous papers \cite{baghdadi2015toward, arpaia2017transport} we have shown that by slowly cooling down (cooling rate 5 $^\circ$C/min) the films after the deposition, at an oxygen pressure of 900 mbar, we promote the full oxidation of the YBCO chains, therefore achieving slightly overdoped films.

To explore the optimally doped and the underdoped regions of the superconducting dome, soon after the deposition each film is cooled down at a post-annealing pressure, lower than the one used to achieve slightly overdoped samples. To ensure the stability of the pressure during the post-annealing, which favors the reproducibility of the results, we have cooled down the films at a constant and continuous oxygen flow: with this procedure we may use pressures down to $1 \cdot 10^{-5}$ mbar, with a stability which is guaranteed within $2 \cdot 10^{-6}$ mbar (see Table  \ref{UDTable} for a list of the postannealing pressures we have used).

Similar experiments are reported in literature: starting from optimally doped films, the oxygen content is reduced ex-situ through  a subsequent annealing at low oxygen pressure \cite{osquiguil1992controlled, tolpygo1996universal, farnan2000fabrication, wuyts1996resistivity}. In our case, we have chosen an in-situ annealing to achieve a better degree of homogeneity and reproducibility of the results.

We have chosen to work with two different substrates to check if the strain is responsible for any change in the doping dependence of the electronic properties of the YBCO films, similarly to what has been previously shown in other HTS cuprates \cite{sato20002, sato2000absence, sato2001influence}. (001) oriented STO substrates are characterized by in plane lattice parameters $a_{\mathrm{STO}} = a_{\mathrm{sub}} = 3.905$ \AA \, slightly larger than those of bulk YBCO, $a_{\mathrm{bulk}} \approx 3.85$ \AA \, (given by the average of the $a$-axis and $b$-axis parameters): the lattice mismatch $\delta^m = 1 - a_{\mathrm{bulk}}/a_{\mathrm{sub}}\approx +2\%$ leads to strained superconducting films, with a very small tensile stress of the  YBCO cell.  On the contrary on (110) oriented MgO ($a_{[0,0,1]} = 4.21$ \AA, $a_{[1,-1,0]} = 5.96$ \AA) a large in-plane film-substrate mismatch is present ($\delta^m \approx +9\%$ and $\approx +35\%$ along the [0,0,1] and [1,-1,0] MgO directions respectively). Even though the in-plane mismatch should induce a tensile strain to the YBCO cells, an in-plane compressive strain has been previously reported on YBCO \cite{broussard1998structural, baghdadi2015toward}, probably related to reconstruction of the MgO surface due to annealing at the high temperatures, required to deposit YBCO \cite{wu1997investigation}. YBCO grows twinned, i.e. with a random exchange of the $a$ and $b$ axis parameters, both on (110) MgO and on (001) STO \cite{schweitzer1996twinning}. This occurs both because of the in-plane symmetries of the substrates, and because of the chosen deposition conditions.

The thickness $t$ has been chosen to be 30 nm: the postannealing becomes more effective at thicknesses $\leq 30$ nm, favoring homogeneous films, with the same oxygen content throughout the thickness.

\section{Study of the temperature dependence of the resistance} \label{sec: rt}

\begin{figure}[hbtp!]
\centering
\includegraphics[width=6.45cm]{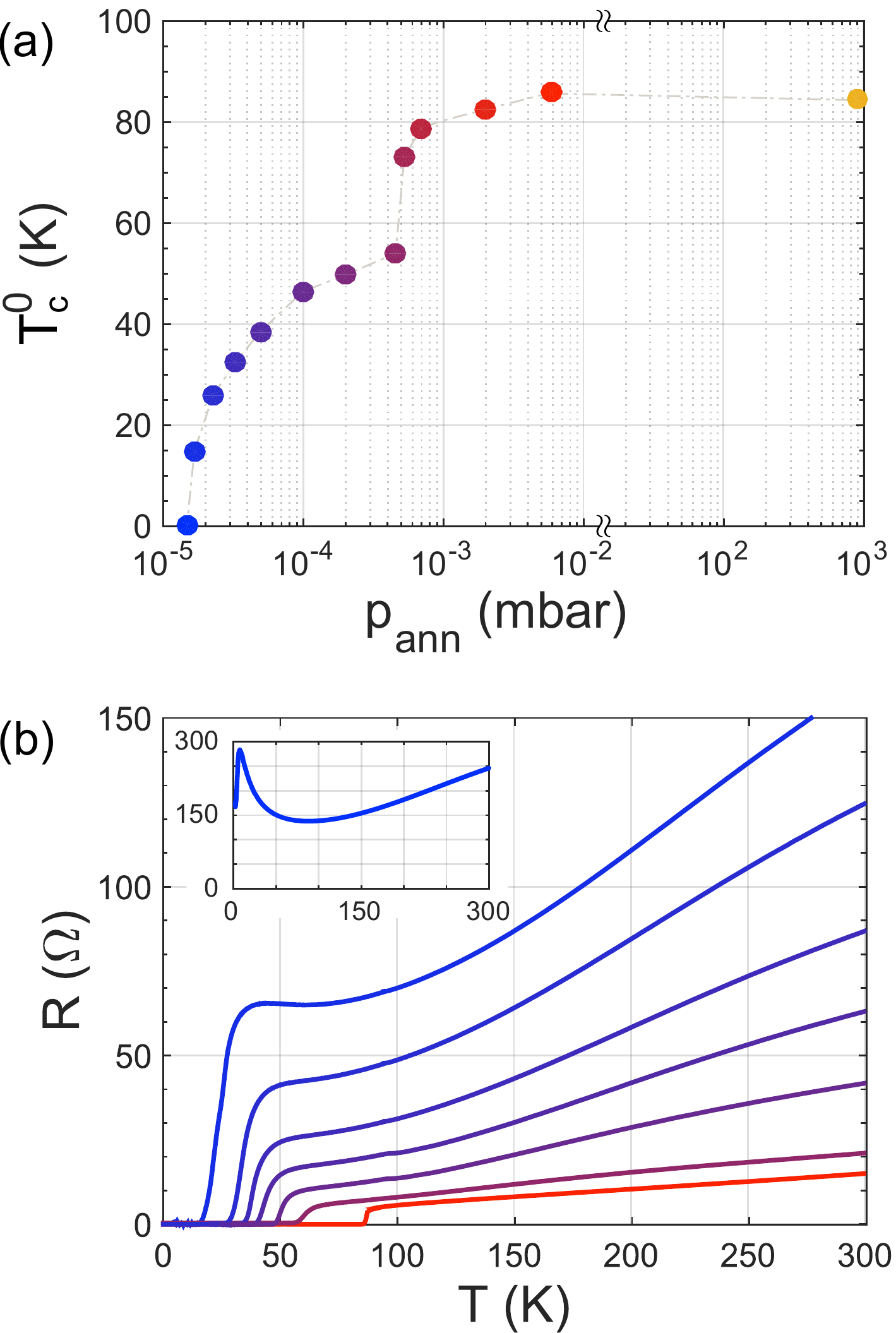}
\caption{(a) Zero resistance critical temperature $T_{\mathrm{c}}^0$ of the 30 nm thick YBCO films on MgO (110) as a function of the post-annealing pressure. (b) The resistance vs temperature $R(T)$ of the YBCO films, with $T_{\mathrm{c}}^0 = 14.9$, $25.8$, $32.3$, $38.5$, $46.4$, $54.0$ and $85.9$ K are presented. In the inset, the $R(T)$ of a strongly de-oxygenated, non-superconducting, YBCO film is also shown.} \label{fig:UD01}
\end{figure}

The resistance vs temperature $R(T)$ of the films has been measured with the Van der Pauw method \cite{van1958method} in the range between 5 and 300 K. By decreasing the postannealing pressure, we have observed a progressive reduction of the zero resistance critical temperature $T_{\mathrm{c}}^0$, as expected for underdoped films, achieving superconducting films with  $T_{\mathrm{c}}^0$ down to $\approx 15$ K (see Fig. \ref{fig:UD01}a, 
\begin{table*}[hbt!] 
\begin{tabular}{c c c c c c c c c c c} 
\hline \hline
$p_{\mathrm{ann}}$ (mbar) &  $T_{\mathrm{c}}^0$ (K) & $\overline{T_{\mathrm{c}}}$ (K) &  $T'$ (K) & $T^{**}$ (K)  & $T^*$ (K) & $R_0$ ($\Omega$)& $R_{\square}$ ($\Omega$) & $\rho_{100 K}$ ($\mu\Omega$cm) & $\Delta T_{\mathrm{c}}$ (K) & $\Delta 2 \theta$ ($^{\circ}$)\\
\hline
$1.5 \cdot 10^{-5}$ & 0 & 0 & 140 & 197 & - & - & 1115 & 1876 & - & 0.39 \\
$1.7 \cdot 10^{-5}$ & 14.9 & 23.7 & 115 & 183 & - & - & 730 & 952 & 8.0 & 0.42 \\
$2.3 \cdot 10^{-5}$ & 25.8 & 33.1 & 102 & 176 & - & - & 567 & 666 & 6.0 & 0.39 \\
$3.3 \cdot 10^{-5}$ & 32.3 & 37.7 & 88 & 167 & - & - & 394 & 422 & 4.5 & 0.42 \\
$5.0 \cdot 10^{-5}$ & 38.5 & 42.5 & 63 & 156 & - & - & 286 & 286 & 5.0 & 0.42 \\
$1.0 \cdot 10^{-4}$ & 46.4 & 49.9 & 64 & 143 & 250 & 7.6 & 190 & 185 & 3.4 & 0.40 \\
$2.0 \cdot 10^{-4}$ & 49.7 & 52.5 & 64 & 142 & 246 & 4.5 & 104 & 103 & 3.1 & 0.41 \\
$4.6 \cdot 10^{-4}$ & 54 & 60.1 & 69 & 125 & 216 & 4.2 & 95 & 102 & 6.0 & 0.43 \\
$5.3 \cdot 10^{-4}$ & 73.1 & 76.3 & - & 116 & 180 & 3.2 & 85 & 100 & 3.9 & 0.40 \\
$7.0 \cdot 10^{-4}$ & 78.8 & 80.6 & - & 108 & 165 & 2.9 & 82 & 100 & 3.7 & 0.45 \\
$2.0 \cdot 10^{-3}$ & 82.5 & 85.9 & - & 112 & 140 & 2.7 & 73 & 95 & 2.9 & 0.42 \\
$6.0 \cdot 10^{-3}$ & 85.9 & 86.9 & - & - & 99 & 1.1 & 68 & 76 & 2.0 & 0.40 \\
$9.0 \cdot 10^{2}$ & 84.5 & 85.4 & - & - & - & 0.8 & 63 & 65 & 1.7 & 0.42 \\
\hline
\end{tabular}
\caption{Summary of the parameters related to 30 nm thick films deposited on MgO (110) substrates. The films have been grown by changing the postannealing oxygen pressure $p_{\mathrm{ann}}$ from 15 nbar up to 900 mbar. From the $R(T)$ of the films at different doping we have extracted the four temperatures $\overline{T_{\mathrm{c}}}$, $T'$, $T^{**}$, $T^*$, the residual resistance $R_0$, as illustrated in Fig. \ref{fig:UD02}, as well as the temperature $T_{\mathrm{c}}^0$ (plotted in Fig. \ref{fig:UD01}a), the broadening of the superconducting transition $\Delta T_{\mathrm{c}}$, the sheet resistance $R_{\square} = R \pi /\ln{2}$ at 300 K and the resistivity $\rho_{100 K} = R t \pi /\ln{2}$ at 100 K. $\Delta T_{\mathrm{c}}$ is defined as the difference between the temperatures $T_1$ and $T_2$ where the resistance has 90\% and 10\% respectively of the normal state value. $\Delta 2 \theta$, defined as the full width at half maximum of the (005) peak in the XRD $2\theta - \omega$ scan, is also listed at different oxygen doping.} \label{UDTable}
\end{table*}
and Table  \ref{UDTable}).


The good homogeneity of the films is proven by a rather narrow - lower than 5 K in most of the cases - transition broadening (see Table  \ref{UDTable}).

In Fig.  \ref{fig:UD01}b the resistance of several films at different oxygen doping is shown as a function of the temperature. When decreasing the oxygen content, the resistance $R$ increases smoothly at first; then it increases much faster in films with $T_{\mathrm{c}}^0 \lesssim 50$ K. This feature has been previously seen both in YBCO single crystals and thin films \cite{PhysRevLett.86.4907, wuyts1996resistivity}, and the threshold, at which such an abrupt change appears ($p \approx 0.1$), has been associated to a sudden change in carrier concentration \cite{veal1991dependence}.

\begin{figure*}[hbtp!]
\centering
\includegraphics[width=15.5cm]{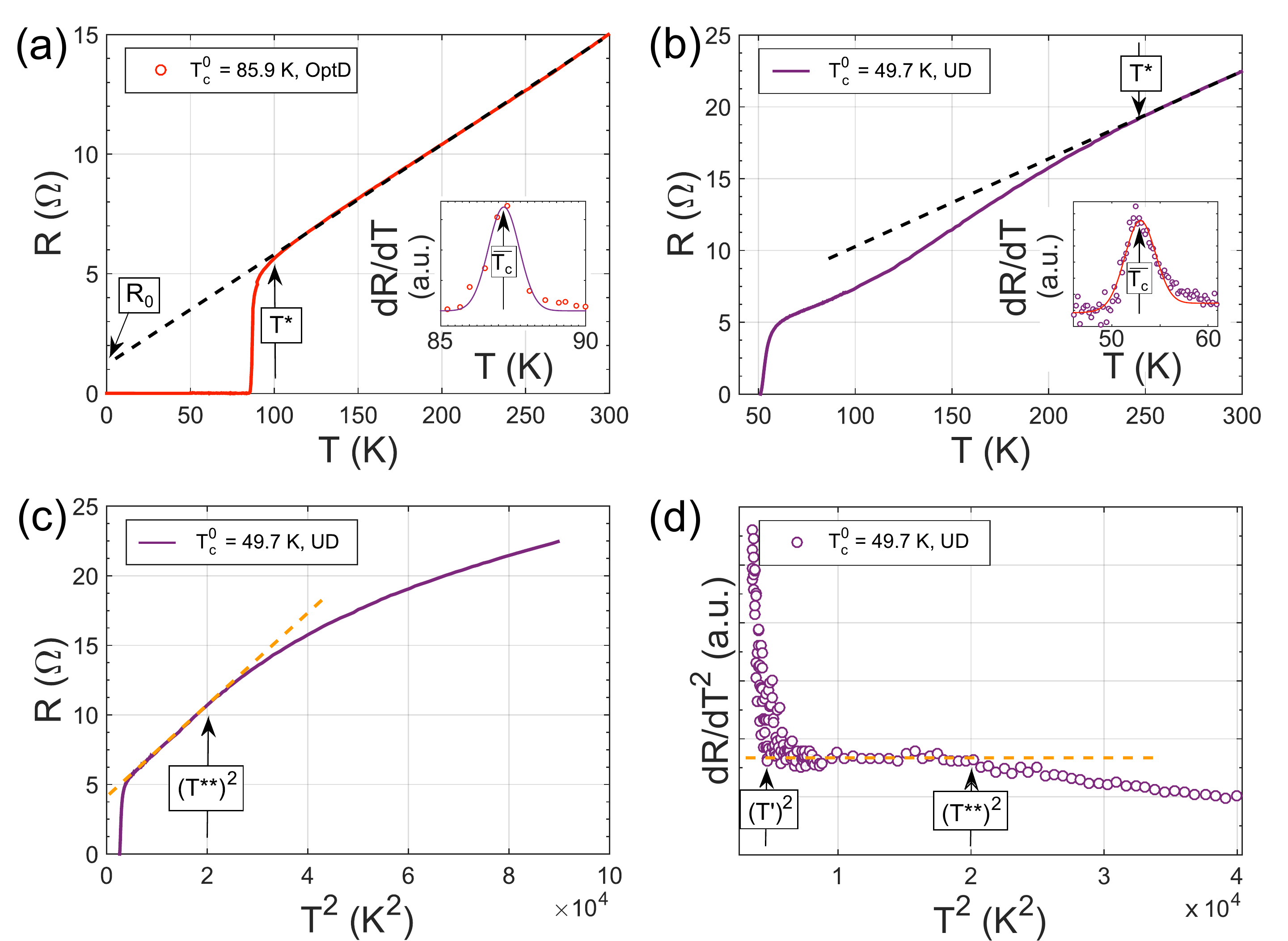}
\caption{The $R(T)$ of a nearly optimal doped film, having a $T_{\mathrm{c}}^0 = 85.9$ K (panel a) and of an underdoped film, having a $T_{\mathrm{c}}^0 = 49.7$ K (panel b), are analyzed and compared. The high temperature resistive behavior has been fitted with the equation $R = R_0 + \epsilon \cdot T$ (dashed line), where the intercept $R_0$ and the slope $\epsilon$ are free parameters. In the nearly optimal doped films, the agreement between the data and the linear fit is very good from 300 K down to 99 K, representing the pseudogap temperature $T^*$. The intercept, $R_0 = 1.1$ $\Omega$, is the residual resistance. For underdoped films, the linear resistive regime is much narrower (in panel b, $T^* = 246$ K) and a quadratic resistive behavior appears at lower temperatures. The residual resistance increases  (in panel b, $R_0 = 4.5$ $\Omega$), while the slope $\epsilon$ is almost independent on the oxygen doping. In strongly underdoped films the linear resistive regime below 300 K is absent, or it is too narrow to be fitted properly, so to unambiguously  determine $T^*$ and $R_0$. In the inset of both panels the $\overline{T_{\mathrm{c}}}$ is extracted as the maximum of the first derivative of the $R(T)$ curve. (c) The resistance of the film shown in panel b is plotted as a function of $T^2$: the linear behavior at low temperature confirms that the dependence of  the $R(T)$ at low temperatures is purely quadratic. (d) First derivative of the $R(T^2)$ data presented in panel c: the temperatures $T' \approx 64$ K and $T^{**} = 142$ K have been extracted respectively from the lower and upper temperatures where $dR/dT^2$ is constant.} \label{fig:UD02}
\end{figure*}

In addition, a clear change in the temperature dependence of the resistance is visible as a function of the doping. The resistance of the films with the highest $T_{\mathrm{c}}^0$ shows the linear temperature dependence from 300 K down to $\approx 100$ K, which is typical of optimally doped films (see red curve in Figs. \ref{fig:UD01}b and \ref{fig:UD02}a). When decreasing the oxygen content, the temperature interval in which $R$ is linear decreases, and a downward bending develops at lower temperatures. In the more de-oxygenated samples, this downward bending is followed, at even lower temperatures, by an upward bending. Finally, in non-superconducting samples this upward bending is even more pronounced, giving rise to a clear metal-insulator transition (see inset of Fig. \ref{fig:UD01}b) \cite{liu2004stripe}.

In underdoped HTS cuprate single crystals, at least four different characteristic temperatures $T^*>T^{**}>T'>T_{\mathrm{c}}$ can be extracted from $R(T)$ curves, acting as crossover lines between different regimes and phases \cite{barivsic2013universal}:

\begin{itemize}
\item[-] {\itshape the pseudogap temperature $T^*$}. It represents the crossover line between the pseudogap region and the still poorly understood strange metal phase.  Since the latter one is characterized by a linear dependence of the resistance on the temperature,  $T^*$ can be inferred by the downward deviation from the $T$-linear behavior.
 \item[-] {\itshape the temperature $T^{**}$}. A crossover from a linear to a purely quadratic resistive behavior can be observed in underdoped films. The temperature  $T^{**}$ represents the upper bound in temperature for the $R \propto T^2$ dependence. This temperature linearly decreases when increasing the doping $p$. It is also close to the characteristic temperatures relative to the maximum of the thermoelectric power \cite{cooper1996some} and the onset of the Kerr rotation signal \cite{kapitulnik2009polar}. Moreover, below $p = 0.1$, this temperature has been recently associated to the onset of the nematicity, an in-plane anisotropy in the transport properties of HTS cuprates, appearing in addition to the anisotropy due to the CuO chains, and whose origin is still debated \cite{cyr2015two}. 
 \item[-] {\itshape the temperature $T'$}. It can be inferred from the $R(T)$ characteristic as the lower bound in temperature for the $R \propto T^2$ dependence. Above $p = 0.1$, it can be associated with the appearance of superconducting fluctuations, which affect the dc conductivity close to $T_{\mathrm{c}}$, in agreement with microwave absorption measurements \cite{grbic2011temperature}. The region of superconducting fluctuations is very narrow, though it expands at low doping, where it can extend to more than 20 K above the superconducting transition. Below $p = 0.1$ it has been related to the metal-insulator crossover temperature \cite{liu2004stripe}.
 \item[-] {\itshape the critical temperature $T_{\mathrm{c}}$}. It represents the crossover line in the $T(p)$ phase diagram below which a superconducting phase appears. Because of the finite broadening of the superconducting transition,  a mean critical temperature $\overline{T_{\mathrm{c}}}$ can be extracted from the maximum of the first derivative of the $R(T)$ characteristic.
\end{itemize} 
The extrapolation of the high-temperature linear behavior of the resistance to zero temperature gives the value of the residual resistance $R_0$. A low value of $R_0$ is the signature of a very clean metallic system, with low disorder. In single crystals it  is almost negligible at the  optimal doping, while it increases when decreasing the doping \cite{tolpygo1996universal, barivsic2013universal}.

The resistance of our films presents the same dependence on the temperature as observed in single crystals, at all oxygen doping. Hence, we can extract the values of  $T^*$, $T^{**}$, $T'$ and $\overline{T_{\mathrm{c}}}$ for our thin films. From the fitting of the high-temperature linear behavior of the resistance we have extracted  the pseudogap temperature $T^*$ and, by extrapolation to zero temperature, the residual resistance $R_0$ (see Figs. \ref{fig:UD02}a and \ref{fig:UD02}b).
Both  the pseudogap temperature  $T^*$ and $R_0$  increase by reducing the oxygen doping. Nearly optimally doped films present a linear dependence of the resistance on the temperature almost down to the critical temperature ($T^* = 99$ K); for underdoped films the interval of linearity decreases. In particular, in films having a $T_{\mathrm{c}}^0$ lower than 46.4 K we do not observe a wide-enough temperature range where the $T$-linear behavior can unambiguously be determined below 300 K, so $T^*$ and $R_0$ cannot be extracted. Underdoped films present instead a purely quadratic resistive behavior at lower temperatures (see Fig. \ref{fig:UD02}c). This Fermi liquid-like dependence of the resistance holds between the two temperatures  $T'$ and $T^{**}$ (see Fig. \ref{fig:UD02}d). The dependence of the temperature $T^{**}$ with the oxygen doping is similar to that of the pseudogap temperature, since $T^{**}$ increases with decreasing $\overline{T_{\mathrm{c}}}$ of the films. Regarding $T'$, it is only few Kelvin higher than $\overline{T_{\mathrm{c}}}$  at all oxygen doping, except for the strongly underdoped films ($p<0.1$), where it increases significantly by reducing the doping. A summary of all the parameters extracted in the most significant samples is listed in Table (\ref{UDTable}). The $R(T)$ of our YBCO thin films as a function of the oxygen doping reproduce the main features of the $R(T)$ of single crystals.

\section{Structural characterization}

The structural properties of the underdoped films have been investigated by Scanning Electron Microscopy (SEM) and X-Ray Diffraction (XRD). 

\begin{figure}[hbtp!]
\centering
\includegraphics[width=8.5cm]{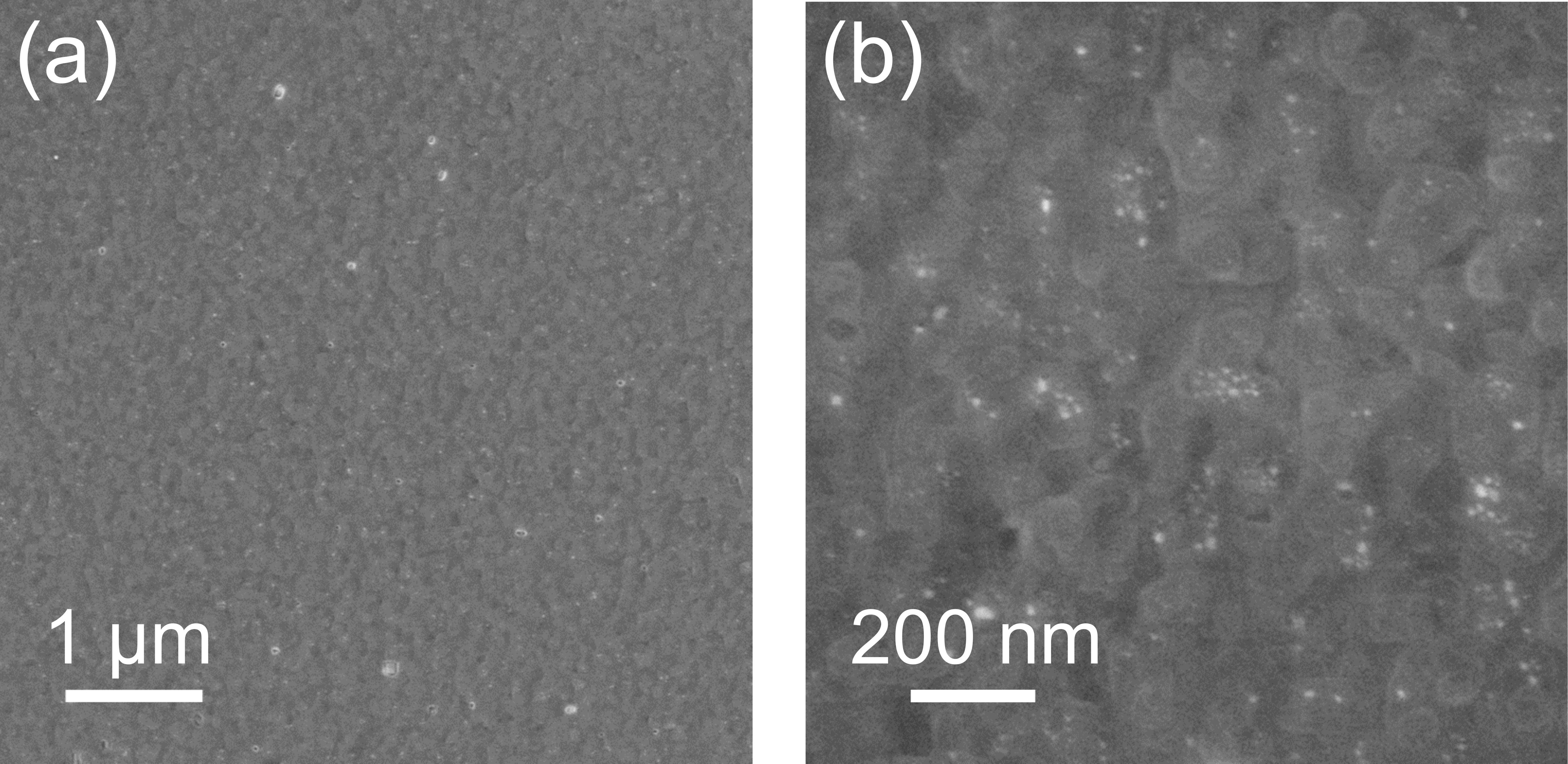}
\caption{(a)-(b) SEM images of a 30 nm thick YBCO film, grown on MgO (110), having $T_{\mathrm{c}}^0 = 49.7$ K.} \label{fig:UD03SEM}
\end{figure}

The surface morphology has been determined by SEM images. The YBCO films present at any doping smooth surfaces, characterized by the typical $c$-axis domains with 3D spirals, and an average roughness which is of the order of one atomic cell (see Fig. \ref{fig:UD03SEM}) \cite{arpaia2017transport}.

\begin{figure}[hbtp!]
\centering
\includegraphics[width=7.7cm]{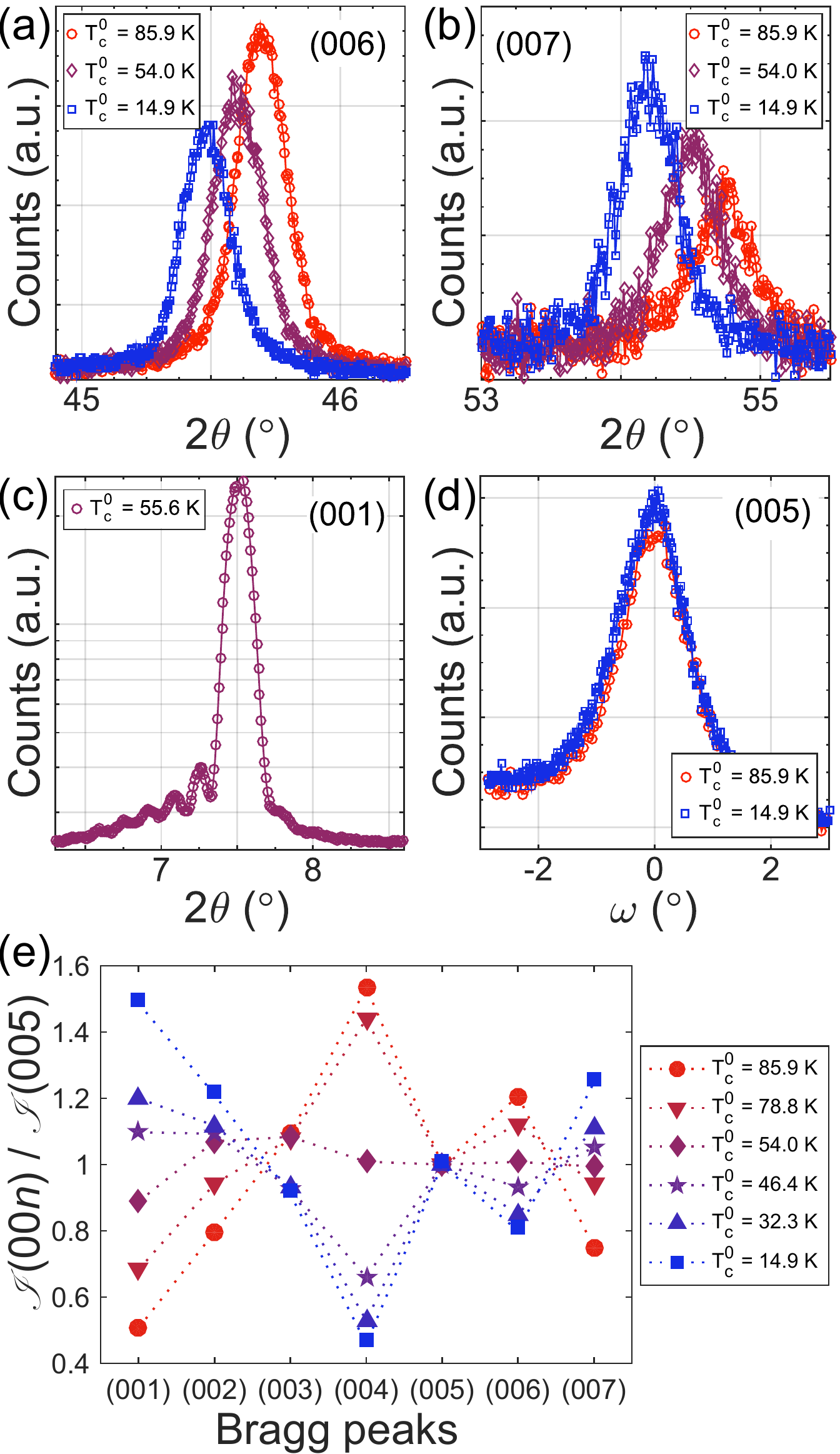}
\caption{XRD analysis on the YBCO films deposited on (110) MgO substrates. (a)-(b) The (006) and (007) Bragg peaks of a nearly optimally doped YBCO film (cirles), of an underdoped film with $T_{\mathrm{c}}^0 = 54.0$ K (diamonds), and of a strongly underdoped film with $T_{\mathrm{c}}^0 = 14.9$ K (squares). When the oxygen content in the films decreases, both peaks shift toward lower angles; the intensity decreases for the (006) peak, while increasing for the (007) one. (c) $2\theta - \omega$ XRD scan of the (001) Bragg reflection of a 50 nm thick YBCO film with $T_{\mathrm{c}}^0 = 55.6$ K. Interference fringes are evident at both sides of the peak. (d) Rocking curves of the (005) peak of the nearly optimal doped sample (circles) and of a strongly underdoped film (squares). The curves are both characterized by the same FWHM of $ \approx 1.4^\circ$, despite the huge difference in $T_{\mathrm{c}}^0$ (of about 70 K) and of $\Delta T_{\mathrm{c}}$ (of about 6 K) of the thin films. (e) Relative change of intensity $\mathscr{I}$ of the (00$n$) Bragg peaks with respect to the (005) Bragg reflection, in several films with different $T_{\mathrm{c}}^0$. For better readability, the $\mathscr{I}$(00$n$)/$\mathscr{I}$(005) values have been divided by 1.3, 0.9, 1.26, 0.06, 1, 0.85, 0.15 for $n = \{1, ..., 7\}$ respectively.} \label{fig:UD03}
\end{figure}

The structural properties have been obtained by XRD analysis. Symmetric $2\theta - \omega$ scans confirm that the films are highly crystalline and $c$-axis oriented (see Fig. \ref{fig:UD03}a and \ref{fig:UD03}b). Interference fringes, which are an indication of high crystallographic quality \cite{arpaia2017transport}, are visible at both sides of the (001) reflection, at all oxygen dopings (see Fig. \ref{fig:UD03}c). 
The full width at half maximum (FWHM) of the $2\theta - \omega$  (0,0,$n$) Bragg peaks, $\Delta2\theta$, which is related to the variation of the $c$-axis parameter within each film, and therefore to the order of the crystal structure, is rather small and independent of the oxygen doping (see Table \ref{UDTable}).

Asymmetrical $2\theta - \omega$ maps around the (038)-(308) YBCO reflections (data not shown) confirmed that the films are twinned on both the substrates. The determination of the two in-plane parameters (see Table \ref{inplanepar}) also led to the confirmation of the strain-state of our films: the YBCO unit cell undergoes an in-plane compressive strain when films are grown on MgO (110) substrates, while the strain is tensile for films grown on STO (001) substrates. 
\begin{table}[!htb]
\begin{tabular}{c|c|c|c}
\hline \hline
\; \; substrate \; \; & \; \; $a$ (\AA) \; \; & \; \; $b$ (\AA) \; \; & \; \; $c$ (\AA) \; \; \\
\hline
MgO (110) & 3.82 & 3.87 & 11.71 \\
STO (001) & 3.84 & 3.90 & 11.67 \\
\hline
\end{tabular}
\caption{\footnotesize{The parameters of the YBCO unit cell, determined via XRD analysis, for fully oxygenated films grown on MgO (110) and  STO (001) substrates. Structural parameters reported for bulk samples with similar oxygen doping ($\delta = 0.07$) are $a = 3.82$ \AA, $b = 3.89$ \AA \, and $c = 11.68$ \AA \, \cite{jorgensen1990structural}.}}
\label{inplanepar}
\end{table}

In the whole range of doping the (0,0,$n$)  rocking curves have the same FWHM (see Fig. \ref{fig:UD03}d). As a consequence, the doping dependent variation of the broadening $\Delta T_{\mathrm{c}}$ of the superconducting transition we have observed cannot be due to structural dishomogeneities or change of crystallinity. In our case $\Delta T_{\mathrm{c}}$ is larger for the more underdoped films, and for films having a  $\overline{T_{\mathrm{c}}} \approx 60$ K (see Table  \ref{UDTable}). This occurrence could be therefore related to intrinsic properties of the YBCO compound at these doping levels, when various local orders are simultaneously at play.

\begin{figure}[hbtp!]
\centering
\includegraphics[width=8.4cm]{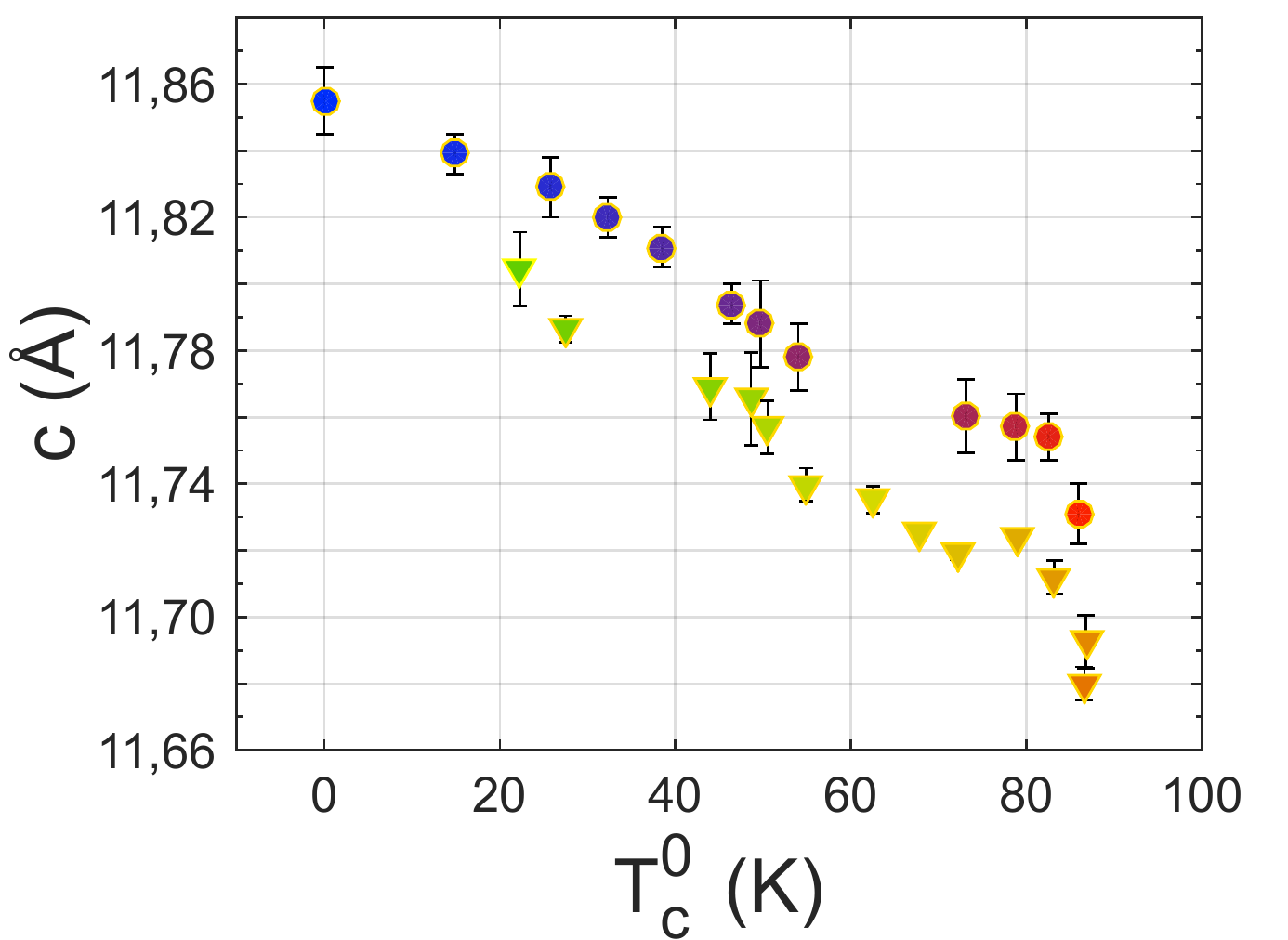}
\caption{The $c$-axis parameter, extracted considering the (00$n$) Bragg peaks in the XRD $2\theta-\omega$ scans,  is shown as a function of $T_{\mathrm{c}}^0$ for both films grown on (110) MgO substrates (circles) and on (001) STO substrates (triangles).} \label{fig:UD03caxis}
\end{figure}

A very significant difference is present in the $2\theta - \omega$ scans measured on films with different $T_{\mathrm{c}}^0$. The reduction of oxygen doping in YBCO is associated to an elongation of the $c$-axis length, which has been observed both in single crystals and in thin films \cite{namgung1988orthorhombic, PhysRevB.48.7554}. In agreement with that, we have observed that the symmetric (00$n$) Bragg peaks in the XRD $2\theta-\omega$ scans appreciably shift toward lower diffraction angles (see Fig. \ref{fig:UD03}a and Fig. \ref{fig:UD03}b).  The extracted $c$-axis lattice parameters of YBCO are shown as a function of the $T_{\mathrm{c}}^0$ in Figure \ref{fig:UD03caxis}, for both films grown on (110) oriented MgO substrates (circles) and on (001) oriented STO substrates (triangles).
In case of films grown on STO, the $c$-axis parameter at a given level of doping is smaller than the bulk value, as a consequence of the tensile strain induced by the substrate into the film. The opposite happens in films grown on MgO, where the substrate - whose surface got reconstructed, once it is heated up to the temperatures required for the deposition of YBCO \cite{wu1997investigation} -  induces a compressive strain and the $c$-axis parameter, at a given oxygen doping, is larger than the bulk value. Apart from the change of the $c$-axis length at a fixed oxygen doping, due to the different film/substrate matching, the $c$($T_{\mathrm{c}}^0$) behavior we got is independent of the used substrate.

Finally, in agreement with previous results on underdoped YBCO thin films \cite{PhysRevB.48.7554}, not only the position, but also the intensity $\mathscr{I}$ of the (00$n$) Bragg peaks changes with $T_{\mathrm{c}}^0$, in opposite directions for different diffraction orders  (see Figs. \ref{fig:UD03}a, \ref{fig:UD03}b and \ref{fig:UD03}e) \footnote{The change of the relative intensities of the (00$n$) Bragg peaks is a direct consequence of the structural change of the YBCO unit cell due to the different oxygen content \cite{jorgensen1990structural}.}.

\section{Doping determination: the thin film phase diagram}

To construct the doping $T$($p$) phase diagram, we have to determine for each film the doping $p$.

\begin{figure*}[hbtp!]
\centering
\includegraphics[width=16.5cm]{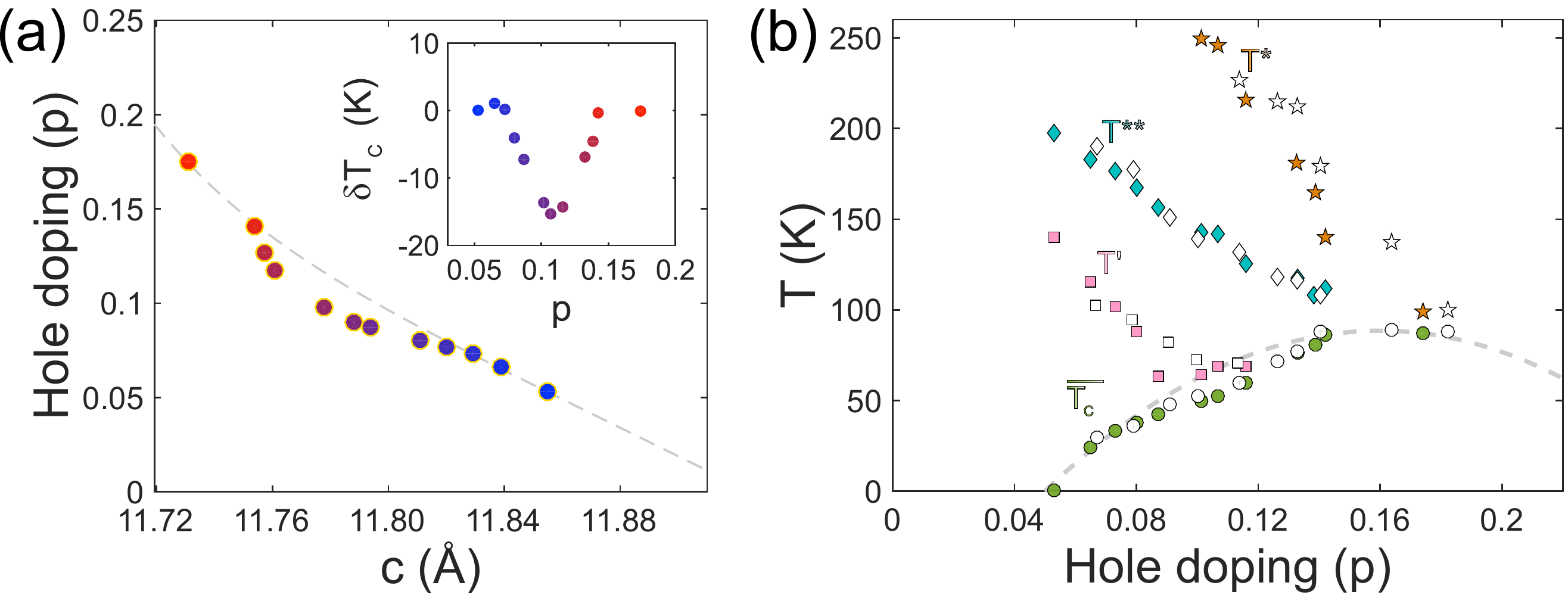}
\caption{(a)  Doping $p$ as a function of the $c$-axis parameter (circles), for the YBCO films grown on (110) MgO substrates. The dashed line represents the  fit to the data at $p < 0.08$ and $p > 0.13$. The inset shows the difference $\delta T_{\mathrm{c}}$ between the measured $\overline{T_{\mathrm{c}}}$ values and the parabolic relation (eq. \ref{eq:  LSCOparab}). The maximum ($\approx$ 18\%) of the $\overline{T_{\mathrm{c}}}$ suppression can be observed corresponding to $p = 0.11$. (b) The four temperatures $\overline{T_{\mathrm{c}}}$, $T'$, $T^{**}$, $T^*$ extracted from the $R(T)$ measurements are plotted as a function of $p$. The resulting phase diagram is very similar to that of YBCO single crystals \cite{barivsic2013universal}. The dashed line is the parabolic relationship (eq. \ref{eq:  LSCOparab}), with $\overline{T_{\mathrm{c}}^{\mathrm{max}}} = 88.5$ K. Filled symbols and empty symbols are related to films grown respectively on (110) MgO and (001) STO substrates.} \label{fig:UD04}
\end{figure*}

The experimental determination of $p$ in thin films of HTS cuprates is extremely difficult, since it requires the knowledge of the oxygen content $n$, which can be accurately determined only in single crystals. Indeed, in bulk crystals chemical and thermogravimetric analyses, implying the possibility of measuring the mass of the oxygen atoms and its variation as a function of the critical temperature,  enable a precise estimate of $n$ \cite{beyers1989oxygen}. 

However, for the specific case of the YBCO compound, the measurement of $p$ is very difficult also for single crystals. As a consequence of the crystal structure, one needs to know how holes are arranged between Cu atoms in the CuO$_2$ planes and Cu atoms in the CuO chains, which depends not only on the oxygen content $n$,  but also on the degree of oxygen ordering in the chains.

Because of the difficulty in determining $p$, the hole doping is commonly calculated in YBCO single crystals from the critical temperature $\overline{T_{\mathrm{c}}}$ by using the empirical parabolic relationship
\begin{equation} \label{eq:  LSCOparab}
1 - \overline{T_{\mathrm{c}}}/\overline{T_{\mathrm{c}}^{\mathrm{max}}} = 82.6 \cdot {(p - 0.16)}^2 \; ,
\end{equation}
where $\overline{T_{\mathrm{c}}^{\mathrm{max}}}$ is the critical temperature at the optimal doping. This equation has been found working well in LSCO, with the exclusion of the data around $p=1/8$ \cite{takagi1989superconductor, presland1991general}. Consequently, it is inaccurate for determining hole doping  around $p=1/8$, and it is not applicable for strongly underdoped, not superconducting, samples.

As shown in Ref. \cite{liang2006evaluation}, the knowledge of the $c$-axis length in single crystals,  in combination with the approximate estimation of $p$ given by  Eq.  \ref{eq:  LSCOparab}, allows to establish a unique relation between the values of $p$ and $c$.

We have applied the same analysis to YBCO thin films. Figure \ref{fig:UD03caxis}  shows  the relation between the $c$-axis parameter and $T_{\mathrm{c}}^0$. To determine the relation between the $c$-axis parameter and $p$, we can substitute in this plot  each $T_{\mathrm{c}}^0$ value with the corresponding $\overline{T_{\mathrm{c}}}$ value (see Table \ref{UDTable}). For every $\overline{T_{\mathrm{c}}}$, the corresponding doping $p$ can be estimated from Eq. \ref{eq:  LSCOparab}, considering $\overline{T_{\mathrm{c}}^{\mathrm{max}}} = 88.5$ K, the value of our nearly optimal doped film. The extracted $p$ as a function of the $c$-axis lattice parameter is  shown in Fig. \ref{fig:UD04}a.

In the interval between $p = 0.05$ and $p = 0.08$, corresponding to YBCO films with 0 K $< \overline{T_{\mathrm{c}}} < 40$ K, the dependence of the doping $p$ on the $c$-axis parameter is linear. The linearity can be extrapolated down to $p=0$, in the strongly underdoped regime.

At intermediate doping, in the interval between $p = 0.08$ and $p = 0.13$, corresponding to YBCO films with 40 K $< \overline{T_{\mathrm{c}}} < 80$ K, the data fall below the low doping linear dependence on the $c$-axis  value. A possible reason for this deviation at intermediate doping ($p \approx 1/8$) is the depression of the critical temperature, associated  to the presence of charge density wave instability \cite{ghiringhelli2012long, chang2012direct}.

From the fit of the data at $p < 0.08$ and $p > 0.13$ (dashed line in Fig. \ref{fig:UD04}a) we can determine the following relationship between $p$ and $c$, which is valid in the entire oxygen content range:
\begin{equation} \label{eq:  pc-eq}
p = 9 y + 1.5 \times 10^9 y^6 \; ,
\end{equation}
where $y = 1 - c/c_0$, and $c_0 = 11.925$ \AA \, is the $c$-axis parameter at zero doping.
The empirical formula  (Eq. \ref{eq:  pc-eq}) is like that found for YBCO single crystals \cite{liang2006evaluation}, but with different prefactors. Therefore our data (the independent measurements of $\overline{T_{\mathrm{c}}}$ and $c$-axis) deviate, similarly to YBCO single crystals, from the  parabolic dependence  which is instead  typically found in YBCO thin films \cite{osquiguil1992controlled}.

The measured $\overline{T_{\mathrm{c}}}$ values can be plotted as a function of the hole doping $p$, obtained using Equation  \ref{eq:  pc-eq}, and compared with those calculated via the parabolic equation (Eq. \ref{eq:  LSCOparab}). The difference $\delta T_{\mathrm{c}}$  between the measured and calculated critical temperatures is negligible, except around $p = 0.11$, where a 16 K ($\approx 18$\%) suppression of critical temperature is observed (see inset of Fig.  \ref{fig:UD04}a). The maximum of the $\overline{T_{\mathrm{c}}}$ suppression is very similar to the $\approx 20$\% occurring in LSCO and YBCO single crystals at $p = 0.125$ \cite{takagi1989superconductor, liang2006evaluation}. 

Finally, one can plot the four temperatures $\overline{T_{\mathrm{c}}}$, $T'$, $T^{**}$, $T^*$ extracted from the $R(T)$ measurements as a function of the hole doping $p$, so to get the $T$($p$) phase diagram of our underdoped YBCO films (see Fig.  \ref{fig:UD04}b). This phase diagram is in full qualitative agreement with that of YBCO single crystals \cite{ando2004electronic, barivsic2013universal, hucker2014competing}. In addition to that, our result seems to exclude that the strain induced by the substrate may have any role in changing the doping dependence of the electronic properties of our twinned YBCO films,  including the depression of the $T_{\mathrm{c}}$ at the $p = 1/8$ doping. Indeed the $T$($p$) data achieved from films deposited both on (110) MgO (filled symbol in Fig. \ref{fig:UD04}b) and on (001) STO (empty symbols in Fig. \ref{fig:UD04}b) substrates fall on top of each other.

\section{Underdoped nanowires}

To realize underdoped nanowires, we have patterned three 50 nm thick Au capped YBCO films grown on MgO (110): the first one is slightly underdoped, with $T_{\mathrm{c}}^0 = 81.4$ K; the second one has a doping close to the 1/8 plateau, with $T_{\mathrm{c}}^0 = 60.5$ K; the third one is strongly underdoped, with $T_{\mathrm{c}}^0 = 42.4$ K.

\begin{figure}[hbtp!]
\centering
\includegraphics[width=7.7cm]{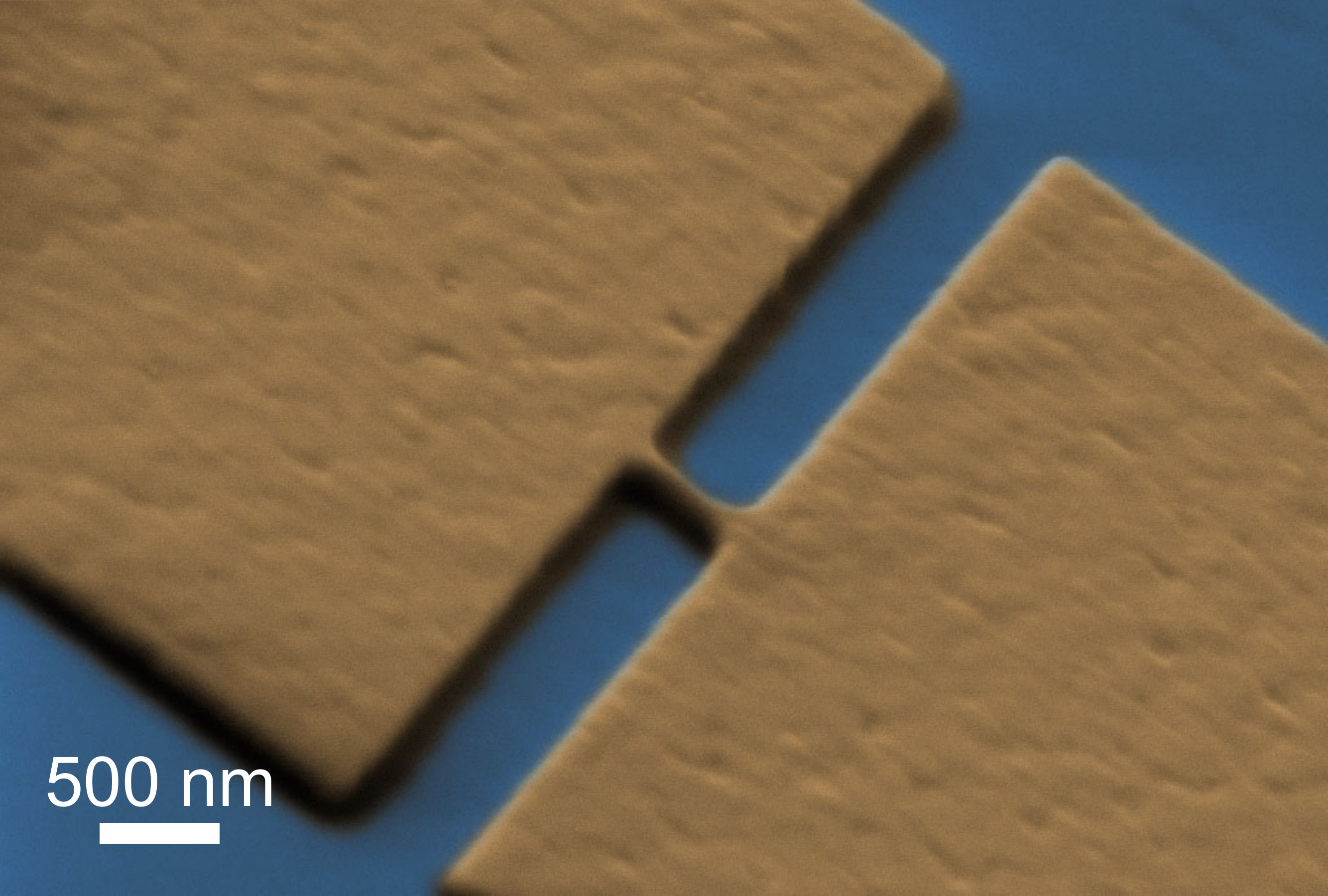}
\caption{SEM image of a 65 nm wide nanowire, patterned on a Au capped underdoped YBCO film, with $T_{\mathrm{c}}^0 = 81.4$ K, grown on a MgO (110) substrate.} \label{fig:SEMwire}
\end{figure}

On each of these films, we have fabricated tens of nanowires, with widths $w$ down to 65 nm and lengths $l = 100$ nm (see Fig. \ref{fig:SEMwire}). For the nanopatterning, we have used the same  procedure, based on an e-beam-lithography-defined carbon mask and a gentle Ar$^+$ ion milling, which we have described in details in our previous works \cite{nawaz2013microwave, arpaia2013improved, nawaz2013approaching}.

Electrical characterization of the underdoped nanowires has been performed via current-voltage characteristics IVCs and resistance versus temperature $R(T)$ measurements. The results of these measurements have been compared with those obtained on slightly overdoped nanowires, having the same geometry, but patterned on a film having $T_{\mathrm{c}}^0 = 85$ K. 

At any doping, all the measured nanowires are superconducting. This is a remarkable result, in particular in the strongly underdoped regime, where previous attempts reported in literature were unsuccessful in preserving superconductivity \cite{Bonetti2008, carillo2012coherent}.

\begin{figure}[hbtp!]
\centering
\includegraphics[width=7.7cm]{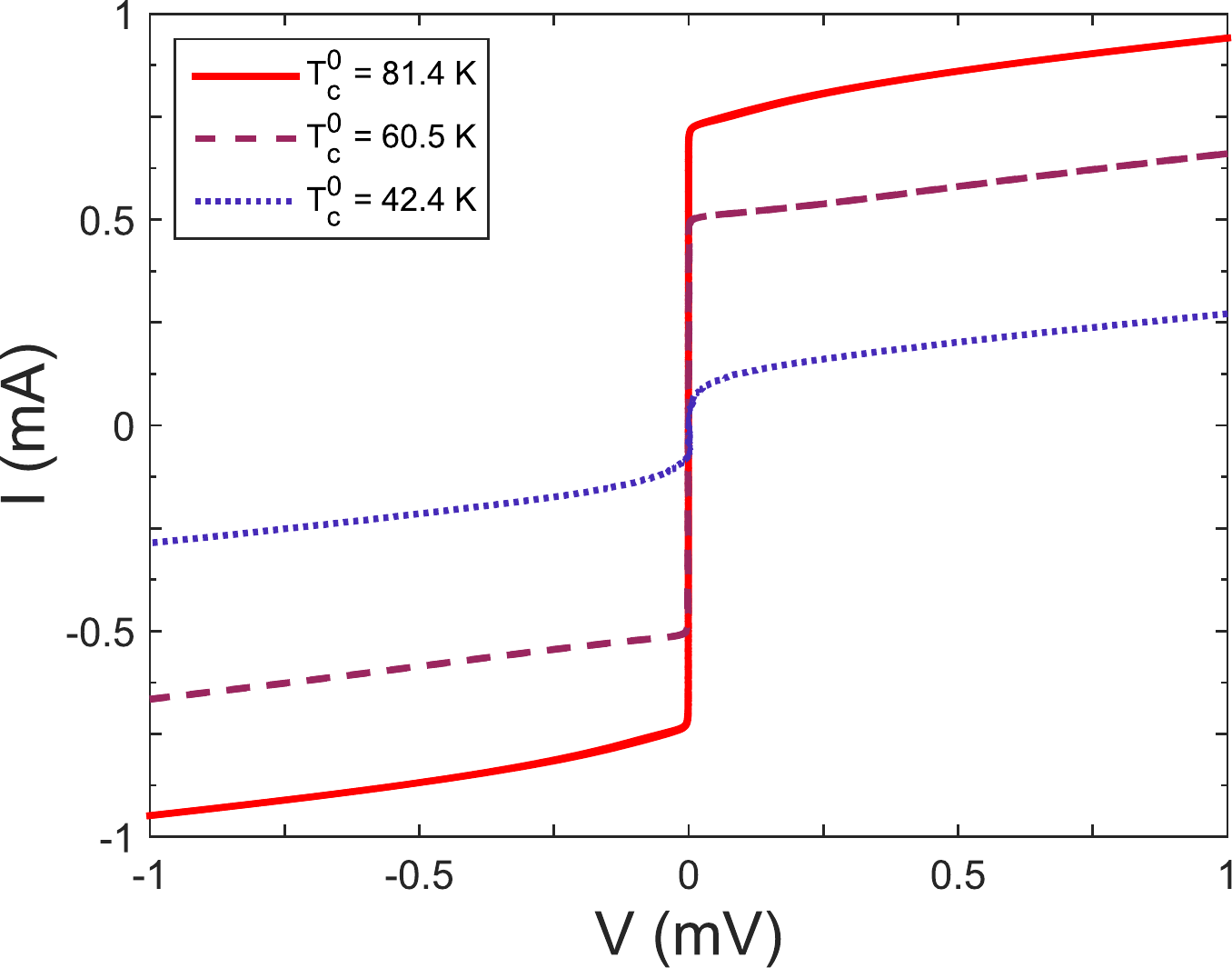}
\caption{$R(T)$ of Au capped YBCO nanowires ($w = 65$ nm), having different oxygen doping ($T_{\mathrm{c}}^0$ is the zero resistance critical temperature of the films, before the patterning). For better readability, the IVCs (both the voltage and the current values) of the nanowires with $T_{\mathrm{c}}^0 = 60.5$ K and  $T_{\mathrm{c}}^0 = 42.4$ K have been multiplied respectively by a factor 1.5 and 10.} \label{fig:IVCwire}
\end{figure}

The IVCs  measured at $T = 4.2$ K, exhibit a typical flux flow like behavior (see Fig. \ref{fig:IVCwire}) \cite{arpaia2014highly}. 
From the IVCs we can determine the critical current $I_{\mathrm{c}}$ of the nanowires, and extract the critical current density $J_{\mathrm{c}} = I_{\mathrm{c}}/(w \cdot t)$. 
For each sample, we can define a $\bar{J}_\mathrm{c}$, which is the average of the $J_{\mathrm{c}}$ values extracted within the wire population.
The $\bar{J}_\mathrm{c}$ values, reported in Table \ref{UDExtrJc}, decrease by two orders of magnitude going from overdoped nanowires to strongly underdoped nanowires. 
\begin{table}[!htb]
\begin{tabular}{c|c|c|c|c|c}
\hline \hline
doping & $T_{\mathrm{c}}^0$ (K) & $\lambda_0$ (nm) & $\xi_0$ (nm) & $J_\mathrm{v}$ (A/cm$^2$) & $\bar{J}_\mathrm{c}$ (A/cm$^2$)\\
\hline
UD & 42.4 & 1000 & 5.0 & $1.7 \cdot 10^6$ & $5.1 \cdot 10^5$ \\
UD & 60.5 & 375 & 5.2 & $1.1 \cdot 10^7$ & $9.9 \cdot 10^6$ \\
UD & 81.4 & 300 & 4.1 & $2.3 \cdot 10^7$ & $2.2 \cdot 10^7$ \\
OD & 85 & 250 & 2.5 & $5.4 \cdot 10^7$ & $5.6 \cdot 10^7$ \\
\hline
\end{tabular}
\caption{\footnotesize{The parameters extracted from underdoped (UD) YBCO nanowires are summarized and compared with those of slightly overdoped (OD) nanowires. For the OD sample, $\lambda_0$ and $\xi_0$ have been extracted from the fits of the $R(T)$, considering a vortex slip model (see Ref. \citenum{arpaia2014resistive}). From these values, we have extrapolated $\lambda_0$ and $\xi_0$ of UD nanowires, considering the typical doping dependence of these two parameters in single crystals \cite{ramshaw2012vortex}. Given these two parameters, for each sample we have calculated $J_\mathrm{v}$, maximum critical current density YBCO nanowires can carry because of vortex entry, using eq. (\ref{eq: J_v}). The $J_\mathrm{v}$ value has been compared with $\bar{J}_\mathrm{c}$, average of the measured $J_{\mathrm{c}}$ values extracted within the wire population. The agreement between the two values is fairly good, at any doping.}}
\label{UDExtrJc}
\end{table}
Such a reduction can be attributed  to the change of the zero temperature values  of the London penetration depth $\lambda_0$ and of the coherence length $\xi_0$, occurring in the underdoped regime. Indeed in 2D nanowires (i.e. characterized, as ours, by dimensions $w, l \gg 4.44\xi_0$), the maximum $J_{\mathrm{c}}$ is given by the entry of Abrikosov vortices, driven by the Lorentz force, overcoming the bridge edge barrier \cite{likharev1979superconducting, papari2014dynamics, arpaia2014yba}. The critical current density due to vortex entry, $J_\mathrm{v}$, can approximately be written as \cite{bulaevskii2011vortex, arpaia2014resistive}:
\begin{equation} \label{eq: J_v}
J_\mathrm{v} \simeq 0.826 J_\mathrm{d} \; ,
\end{equation}
where $J_\mathrm{d}$ is the Ginzburg-Landau depairing critical current density \cite{tinkham1996introduction}
\begin{equation}
J_\mathrm{d} = \frac {\Phi_0}{ 3^{3/2}\pi\mu_0 \lambda_0^2\xi_0 } \label{Eq:J_d} \; .
\end{equation}
In the latter equation, $J_\mathrm{d}$ depends on $\lambda_0$ and $\xi_0$, while $\Phi_0$ is the flux quantum and $\mu_0$ the vacuum permeability.

In underdoped nanowires, we cannot extract $\lambda_0$ and $\xi_0$ from the $R(T)$,  as we have done on slightly overdoped nanowires \cite{arpaia2013improved, arpaia2014resistive}, for reasons which will be clarified later in this section.
In single crystals, the dependence of the London penetration depth and of the coherence length on the oxygen doping has been determined by electron-spin resonance measurements and by the values of the upper critical magnetic field respectively \cite{pereg2004absolute, ramshaw2012vortex}. We can therefore estimate $\lambda_0$ and $\xi_0$ in our patterned underdoped nanowires, assuming 1) that the values of $\lambda_0$ and $\xi_0$ at the optimally doped regime are those we have extracted from the fits of the  $R(T)$ of our slightly overdoped nanowires, considering vortex slip model \cite{arpaia2014resistive}, and 2) that the dependence of the London penetration depth and of the coherence length on the oxygen doping is the same in single crystals and our nanostructures.

By knowing $\lambda_0$ and $\xi_0$, for each underdoped  film we can calculate, using Eq. (\ref{eq: J_v}), the maximum critical current density that the YBCO nanowires can carry because of vortex entry, $J_\mathrm{v}$, and compare this value with the measured $\bar{J}_c$. A summary of all the measured and calculated parameters for the different films at various doping is enclosed in Table \ref{UDExtrJc}; the agreement between $J_\mathrm{v}$ and $\bar{J}_c$ is fairly good.

\begin{figure}[bhtp!]
\includegraphics[width=7cm]{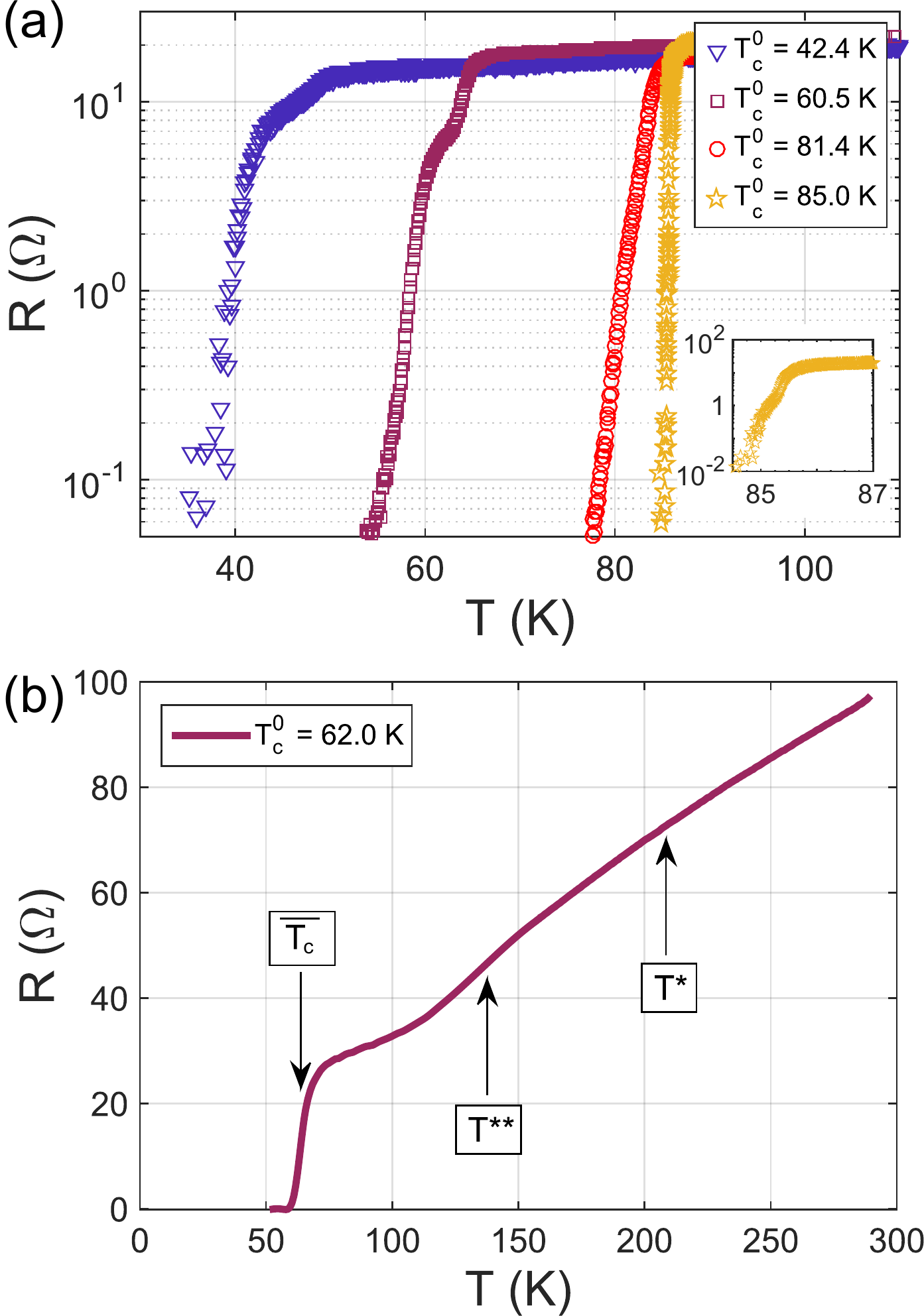}\\
\caption{(a) $R(T)$ of Au capped YBCO nanowires ($w = 140$ nm), having different oxygen doping ($T_{\mathrm{c}}^0$ is the zero resistance critical temperature of the films, before the patterning). For clarity, the $R(T)$ of the slightly overdoped nanowire (star symbols), used for comparison, is presented in the inset in a narrower temperature range. (b) $R(T)$ of a bare nanowire ($w = 100$ nm), patterned on an underdoped YBCO film with $T_{\mathrm{c}}^0 = 62.0$ K. The characteristic temperatures $\overline{T_{\mathrm{c}}}$, $T^{**}$ and $T^{*}$ have been extracted following the procedure described in Section \ref{sec: rt} and their values are in agreement with those reported in Table \ref{UDTable}
} \label{Fig: UDNW}
\end{figure}

A further confirmation of the quality of the nanowires with different oxygen doping comes from the measurements of the 
resistance vs temperature $R(T)$, shown in Fig. \ref{Fig: UDNW}a. 
Here, in the geometry we have chosen (see Fig. \ref{fig:SEMwire}), with the nanowire embedded between two wide electrodes, we see both the superconducting transition of the electrodes, which mostly coincides with that of the as-grown film, and the transition of the nanowire, which is broader and occurs at lower temperature.
It is evident that the onset temperature of the superconducting transition related to the nanowire is only a few Kelvin lower than the one of the wide electrodes, independently of the oxygen doping. This circumstance never occurred in previous works on underdoped HTS nanostructures.


The broadening of the superconducting transition associated to the nanowire increases at lower doping, as expected considering dissipation induced by Abrikosov vortices crossing the nanowires \cite{arpaia2014resistive}, as a consequence of the larger values of $\lambda_0$ and $\xi_0$. However, at the moment a quantitative analysis of this transition is still missing. Indeed, the broadening of the transition of the unpatterned underdoped films, which is related to intrinsic properties of the YBCO compound at specific doping levels, influences the transition broadening of the nanowires, possibly giving rise to a substantial contribution which adds to that coming from the entry of Abrikosov vortices.

From the measurement of the critical current density and of the critical temperature, we conclude that the superconducting properties of Au capped YBCO nanowires are unaffected by the nanopatterning procedure. However the electronic properties in the normal state are hidden, since $R_{\square}$ is dominated by the Au capping layer, acting as an electrical shunt \cite{arpaia2013improved}. To study the physics of the YBCO nanostructures at temperatures larger than the superconducting transition temperature, we have fabricated bare underdoped YBCO nanowires, which are protected during the nanopatterning procedure only by a hard carbon mask, deposited by pulsed laser deposition at room temperature and removed by an oxygen plasma etching at the end of the fabrication. The fabrication of nanowires, unprotected by gold, is slightly detrimental to the superconducting properties \footnote {The reason lays in the thermal conductivity of carbon, which is - differently than Au - much lower than that of YBCO and it cannot prevent the overheating of the superconducting layer during the baking of the resists and the Ar$^+$ ion milling.}, as we have already investigated in the slightly overdoped region \cite{arpaia2016improved, arpaia2017transport}. However, from the $R(T)$ of these nanowires we can extract, following the procedure described in Section \ref{sec: rt} for thin films, all the temperatures characterizing the phase diagram of YBCO in the normal state (see Fig. \ref{Fig: UDNW}b). 

\section{Summary and conclusions}
We have succeeded in fabricating and studying high quality YBCO thin films as a function of the oxygen doping, from the slightly overdoped regime down to the strongly underdoped region. The phase diagram we can build, from the analysis of the $R(T)$ of the films, has strong analogies with that of YBCO single crystals, both in the superconducting and in the normal state. In particular, the superconducting dome of our thin films is depressed in correspondence of the $p = 1/8$ doping, which is a 
feature suggestive of a strong competition of superconductivity with the CDW order \cite{ghiringhelli2012long, chang2012direct}. The presence of this feature, and more in general the doping dependence on the temperature, characterizing the normal state of the YBCO phase diagram, 
are not strain dependent: we got identical features on twinned films both deposited on  (110) oriented MgO substrates and on (001) oriented STO substrates, where the strain induced into the superconducting layer is respectively compressive and tensile.

We have patterned the thin films into nanowires, both with and without the protection of a Au capping layer. The nanowires,  with widths down to 65 nm, are superconducting at any oxygen doping under investigation and preserve essentially the same properties as unstructured films. Nanowires capped with Au are ideal structures to study the superconducting properties at the nanoscale, since they are characterized by values of the critical current densities, close to the depairing values, and of the critical temperatures, close to the as-grown films. The electronic properties in  the normal state of the YBCO phase diagram can also be studied in bare nanowires: the $R(T)$ dependence of uncapped YBCO nanowires presents indeed signature of all the characteristic temperatures, acting as crossover lines between different regimes and phases. 

Our  underdoped YBCO nanowires pave the way for several experiments aiming at elucidating the connection between superconductivity and the various nanoscale orders playing a role in the underdoped region of the HTS phase diagram.
Ultimately this could lead to a better understanding of the microscopic mechanisms responsible for High critical Temperature Superconductivity.

\begin{acknowledgements}
This work has been supported by the Swedish Research Council (VR) and by the Knut and Alice Wallenberg Foundation. The authors would also like to thank T. Claeson for inspiring discussions.
\end{acknowledgements}
\bibliography{biblio}

\end{document}